\documentclass[aps,prl,twocolumn,superscriptaddress,amsmath]{revtex4-2}
\usepackage{graphicx}
\usepackage{hyperref}
\usepackage{mathrsfs}
\usepackage{bm}
\usepackage{color}
\usepackage{siunitx}
\usepackage[capitalize]{cleveref}
\usepackage{orcidlink}

\hypersetup{hypertex=true,
	colorlinks=true,
	anchorcolor=blue,
	linkcolor=blue,
    citecolor=blue,
	urlcolor=blue}
\begin{document}

\title{Anti-higher-order Weyl semimetal}

\author{Cheng-Ming Miao}
\affiliation{International Center for Quantum Materials, School of Physics, Peking University, Beijing 100871, China}

\author{Yu-Hao Wan}
\email[]{wanyh@stu.pku.edu.cn}
\affiliation{International Center for Quantum Materials, School of Physics, Peking University, Beijing 100871, China}

\author{Qing-Feng Sun}
\email[]{sunqf@pku.edu.cn}
\affiliation{International Center for Quantum Materials, School of Physics, Peking University, Beijing 100871, China}
\affiliation{Hefei National Laboratory, Hefei 230088, China}

\begin{abstract}
Higher-order topology extends the bulk-boundary correspondence by enabling corner or hinge localized states. Here we identify an anti-higher-order Weyl semimetal, a three dimensional topological phase in which the conventional boundary hierarchy is reversed. Unlike conventional higher-order Weyl semimetals where bulk topology enforces hinge Fermi arcs, this phase hosts anti-hinge states, meaning the bulk topology forces states to vanish at specific hinge orientations.
Using a minimal two-band model, we show how Weyl points separate the Brillouin zone into quantum anomalous Hall and anti-higher-order topological insulating regions, with the latter characterized by a band-inversion surface enclosing two distinct high-symmetry points carrying opposite topological charges. Analytical solutions for Weyl points, Berry curvature monopoles, and slice Chern numbers are derived, with numerical simulations confirming the resulting anti-hinge behavior. Our work establishes anti-higher-order topology as a dual counterpart to conventional higher-order phenomena, further extending the exploration of topological phases.
\end{abstract}

\maketitle

\emph{Introduction}.---The discovery of Weyl semimetals has revolutionized our understanding of gapless topological phases, linking bulk Weyl points to surface Fermi arcs through the bulk-boundary correspondence\cite{
wan_topological_2011,lv_experimental_2015,soluyanov_typeii_2015,xu_discovery_2015}. A useful perspective is to view a three-dimensional (3D) Weyl semimetal as a momentum-space stacking of two-dimensional (2D) insulating slices parameterized by a conserved momentum, such that the Chern number of the two-dimensional subsystem changes across the Weyl points \cite{jiang_signature_2017,noh_experimental_2017,ge_experimental_2018,armitage_weyl_2018}. This picture naturally accounts for the emergence of 2D surface states in the form of open Fermi arcs connecting the surface projections of Weyl points with opposite chiralities, a hallmark feature that has been widely observed in experiments and emulated in artificial platforms \cite{huang_weyl_2015,xu_spin_2016,xu_distinct_2017}. Beyond conventional boundary responses, the notion of higher-order topology has recently been generalized to Weyl systems, leading to the concept of higher-order Weyl semimetals (HOWSM) \cite{benalcazar_quantized_2017, schindler_higherorder_2018b,ghorashi_higherorder_2020,wang_higherorder_2020}.
In this standard paradigm, bulk Weyl points give rise not only to 2D Fermi-arc surface states but also to one-dimensional gapless hinge states. These one-dimensional hinge states are widely regarded as the defining fingerprint of higher-order Weyl topology \cite{ghosh_hingemode_2022, zhou_entanglement_2023, rafi-ul-islam_chiral_2024}. Since proposed, HOWSM have been explored in diverse physical contexts, including Floquet extensions \cite{zhu_floquet_2021, wan_floquet_2026}, non-Hermitian counterparts \cite{ghorashi_nonhermitian_2021a,liu_higherorder_2021, bid_nonhermitian_2023}, square-root constructions \cite{song_squareroot_2022}, valley-selective realizations \cite{xiong_valley_2023}, and disorder-driven regimes \cite{shang_disorderinduced_2025}.
Benefiting from their high tunability and controllable symmetries, classical platforms such as acoustic systems \cite{wei_higherorder_2021, pu_acoustic_2023}, photonic metamaterials \cite{pan_real_2023, qi_realization_2025}, and electrical circuits \cite{rafi-ul-islam_chiral_2024, zheng_topolectrical_2022, zheng_exploring_2022} have provided powerful platforms for realizing and experimentally probing HOWSM. Together, these works establish HOWSM as a mature and rapidly expanding class of topological phases across a broad range of physical platforms.

\begin{figure}
  \centering
  \includegraphics[width=0.9\columnwidth,angle=0]{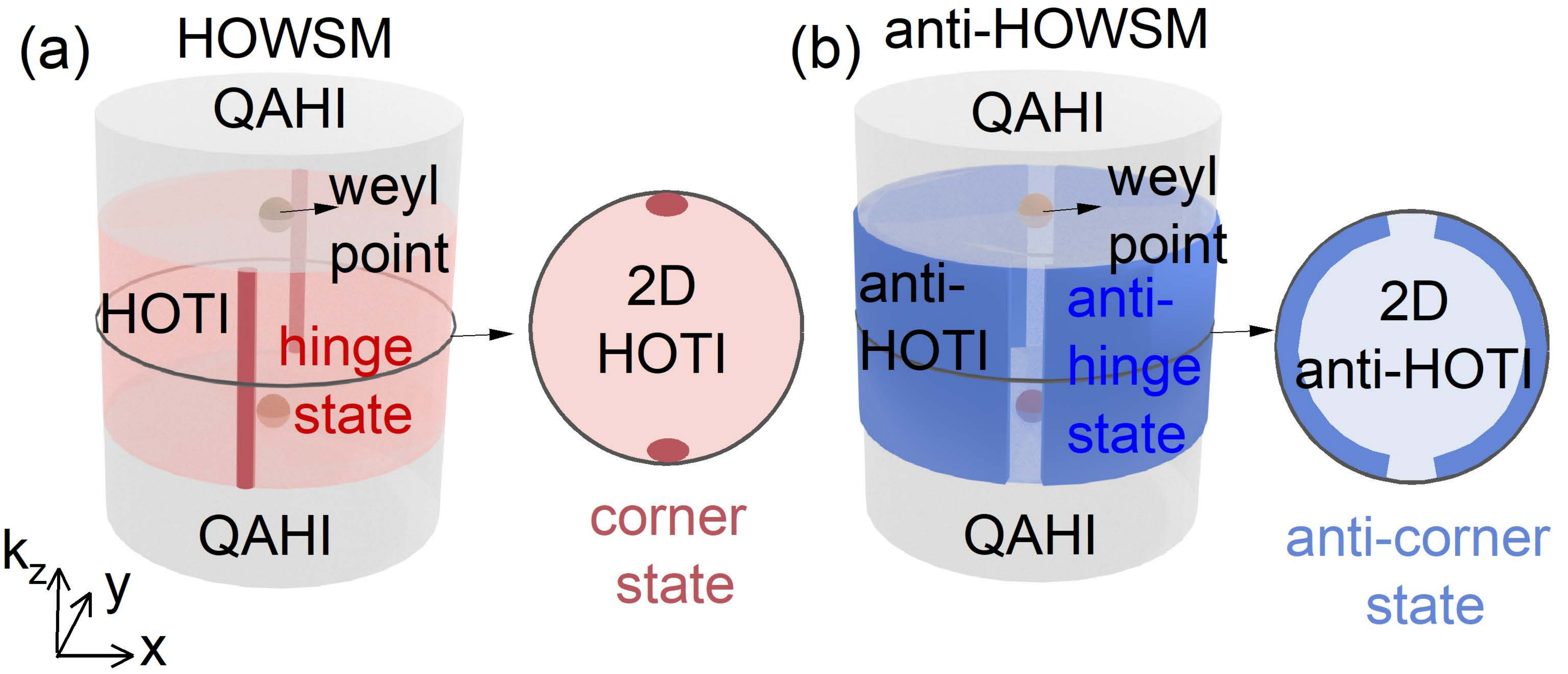}
  \caption{(a) Schematic of a 3D HOWSM (left) and its $k_z=0$ slice on the $x-y$ plane, showing a 2D HOTI phase (right).
  Orange spheres denote Weyl points separating the Brillouin zone into QAHI and HOTI regions. Red lines and ellipses indicate hinge and corner states. (b) Schematic of an anti-HOWSM (left) and its $k_z=0$ slice on the $x-y$ plane, showing a 2D anti-HOTI phase (right).
  Orange spheres denote Weyl points separating QAHI and anti-HOTI regions. Blue incomplete surfaces and broken circles indicate the absence of hinge/corner states, corresponding to anti-hinge and anti-corner states.}
  \label{fig1}
\end{figure}

In this Letter, we introduce a topological phase that defies this additive logic: the anti-HOWSM. As illustrated in Fig. \ref{fig1}(a), a conventional HOWSM is characterized by bulk Weyl points that separate the 3D Brillouin zone into momentum-space regions realizing either quantum anomalous Hall insulator (QAHI) or higher-order topological insulator (HOTI) phases. This structure gives rise to the coexistence of 2D Fermi-arc surface states and robust one-dimensional hinge modes, reflecting the standard higher-order bulk–boundary correspondence. 
In contrast, the anti-HOWSM exhibits an entirely inverted boundary response [Fig. \ref{fig1}(b)]. Here, bulk Weyl points partition the Brillouin zone into QAHI and anti-HOTI momentum slices. Despite the presence of gapless surface states on most surface orientations, one-dimensional hinge states are selectively absent, defining an anti-hinge response. From the momentum-space slicing perspective, this behavior originates from 2D anti-HOTI phases, in which boundary states persist along edges but are systematically excluded from particular corner locations. As a result, the anti-HOWSM realizes a boundary hierarchy that inverts the conventional higher-order picture, suppressing concentrating boundary modes at lower-dimensional boundaries.
We construct a minimal two-band model to realize this phase, providing a phase diagram that maps the interplay between bulk Weyl points and these exotic subtractive boundary signatures.
Our work expands the horizon of Weyl physics, showing that topological protection can enforce the absence, rather than just the presence, of electronic states, thereby elucidating the significance of anti-topology as the dual counterpart to the conventional topological paradigm.

\emph{Two bands model}.---We start from a simple 3D two bands model for anti-HOWSM: $H(\mathbf{k})={\mathbf{d}(\mathbf{k}) \cdot \bm{\sigma}}$,
where $\mathbf{k} = (k_x, k_y, k_z)$ is a wave vector in the first Brillouin zone. $\mathbf{d}(\mathbf{k})=[d_x(\mathbf{k}),d_y(\mathbf{k}),d_z(\mathbf{k})]$ is a vector with three components being given functions of $\mathbf{k}$ and $\bm{\sigma}=(\sigma_{x}, \sigma_{y}, \sigma_{z})$ are the Pauli matrices acting on the orbit/spin space.
The representations for the components of the vector $\mathbf{d}(\mathbf{k})$ are given below:
\begin{align}   d_x(\mathbf{k})&=\sin{k_x}, \qquad \qquad d_y(\mathbf{k})=\sin{k_y},\nonumber  \\
d_z(\mathbf{k})&=m+B_x \cos{k_x}+B_y\cos{k_y}+B_z\cos{k_z},
  \label{eq2}
\end{align}
where $m$ is the mass parameters and $B_{x/y/z}$ is the hopping parameters for $x/y/z$ direactions. It is well established that realizing Weyl points necessitates breaking either time-reversal symmetry ($\mathcal{T}$) or inversion symmetry ($\mathcal{P}$) \cite{wan_topological_2011}. Our model breaks $\mathcal{T}$ while explicitly preserving inversion symmetry. The inversion operator is given by $\mathcal{P} = \sigma_z$, satisfying $\mathcal{P} H(\mathbf{k}) \mathcal{P}^{-1} = H(-\mathbf{k})$. By contrast, the Hamiltonian does not satisfy time-reversal symmetry under either the spinless form $\mathcal{T}=K$ or the spinful form $\mathcal{T}=i\sigma_yK$.
A $\mathcal P$-broken but $\mathcal T$-preserved realization can also be constructed in the enlarged Kramers degenerate models. 
A theoretical guideline for such a construction is provided in Sec.~S-1 of the Supplemental Materials (SM) \cite{seesupplemental}. In the following, we focus on the minimal $\mathcal P$-preserved and $\mathcal T$-broken model in Eq.~(\ref{eq2}). 
To explore the phase diagram, we introduce an anisotropy parameter $\lambda$ by setting $B_y = \lambda B_x$ (with $\lambda \neq 1$). This choice breaks the $C_4$ rotational symmetry around the $z$-axis while maintaining $\mathcal{P}$, providing a tunable knob to manipulate the interplay between bulk Weyl points and the subtractive boundary signatures.

The energy dispersion of the two‑band Hamiltonian is given by $E_{\pm}(\mathbf{k}) = \pm \sqrt{d_x^2(\mathbf{k})+d_y^2(\mathbf{k})+d_z^2(\mathbf{k})}$.
Weyl points occur when \(E=0\). This requires the three conditions: $d_x(\mathbf{k}) = 0, d_y(\mathbf{k}) = 0, d_z(\mathbf{k}) = 0$. The first two equations force \(k^0_x = 0\) or \(\pm \pi\) and \(k^0_y = 0\) or \(\pm \pi\). Hence the possible in‑plane momenta are the four high‑symmetry points:
\((k^0_x,k^0_y) \in \{(0,0), (0,\pm \pi), (\pm \pi,0), (\pm \pi,\pm \pi)\}\).
For each such pair \((\eta_x,\eta_y)\) with \(\eta_x = \cos k_x = \pm 1\) and \(\eta_y = \cos k_y = \pm 1\), the third equation becomes
\begin{align}
m + B_x \eta_x + \lambda B_x \eta_y + B_z \cos k_z = 0,
  \label{eq4}
 \end{align}
which yields
$
\cos k^0_z = -\,\frac{m + B_x \eta_x + \lambda B_x \eta_y}{B_z}.
$
A pair of Weyl points exists if the right‑hand side lies in the interval \([-1,1]\); then \(k_z\) takes two values \(\pm k_z^{0}\) with $
k_z^{0} = \arccos\!\Bigl[-\,\frac{m + B_x \eta_x + \lambda B_x \eta_y}{B_z}\Bigr]$.
Consequently, the Brillouin zone contains up to Weyl points located at \((k_x,k_y,k_z) = (k_x^{0},k_y^{0},\pm k_z^{0})\).

To determine the chirality of the Weyl points, we can calculate the low-energy Berry curvature as \cite{thouless_quantized_1982,kohmoto_topological_1985}
\begin{align}
\boldsymbol{\Omega}(\mathbf{k}) = \frac{1}{2}\, \frac{\mathbf{d}(\mathbf{k})}{|\mathbf{d}(\mathbf{k})|^3} \cdot \bigl[ \partial_{k_x}\mathbf{d}(\mathbf{k}) \times \partial_{k_y}\mathbf{d}(\mathbf{k}) \bigr].
\end{align}
At the Weyl points, the Berry curvature exhibits a singularity, and its sign determines the chirality. Based on the sign of the Berry curvature, we can determine the chirality of the Weyl points: if the Berry curvature points outward, the chirality is positive; if it points inward, the chirality is negative.
The Chern number for a 2D slice at fixed \(k_z\) is obtained via the integral of the Berry curvature over the \((k_x, k_y)\) Brillouin zone \cite{thouless_quantized_1982,kohmoto_topological_1985}:
\begin{align}
C = \frac{1}{2\pi} \iint_{\text{BZ}} \Omega_{xy}(\mathbf{k}) \, dk_x \, dk_y,
\end{align}
where \(\Omega_{xy}(\mathbf{k})\) is the \(z\)-component of the Berry curvature \(\boldsymbol{\Omega}(\mathbf{k})\).

\begin{figure}
	\centering
	\includegraphics[width=\columnwidth,angle=0]{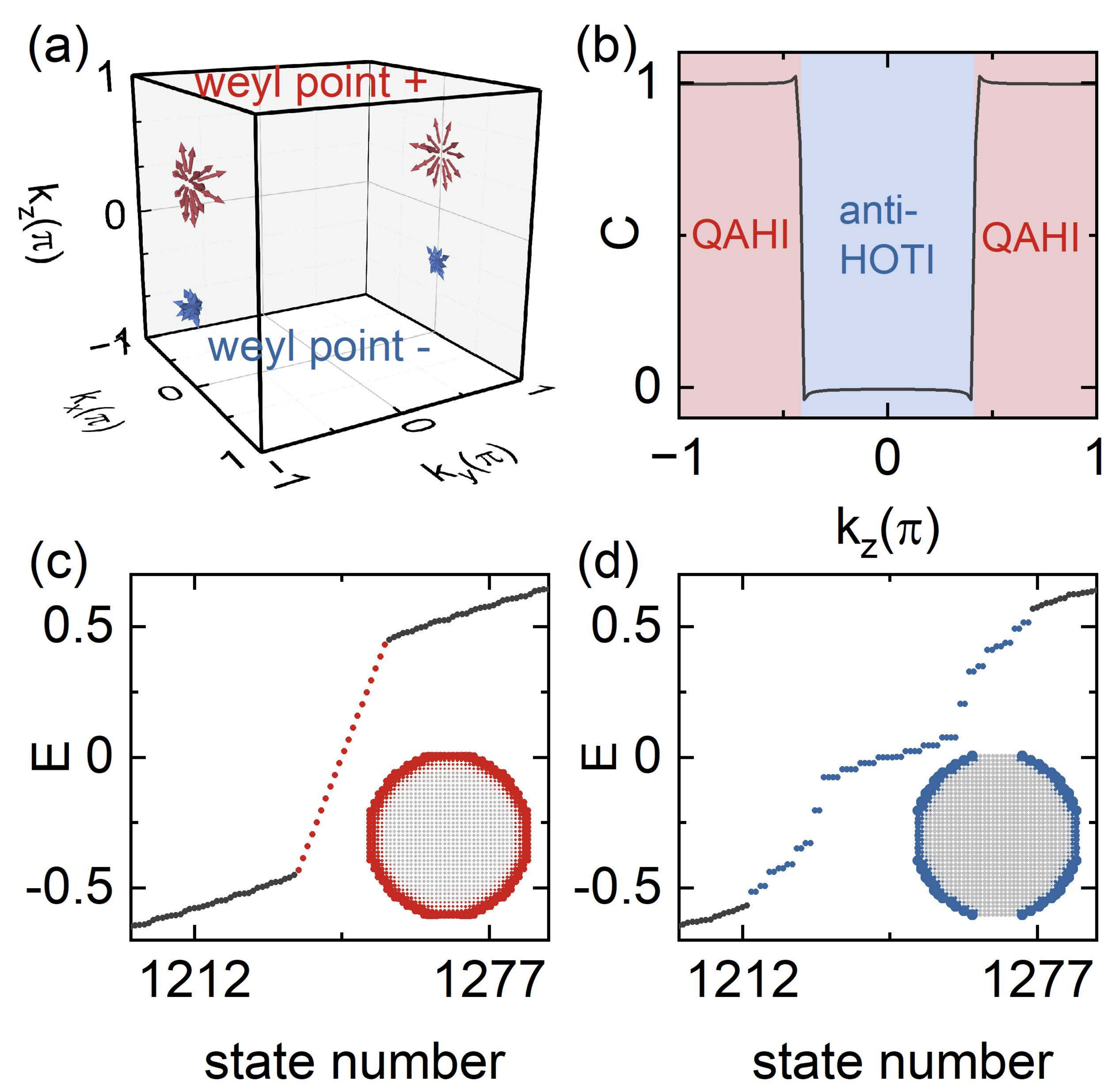}
	\caption{(a) The vector plot of the Berry curvature in momentum space, with only vectors of large magnitude shown. Red (blue) arrows indicate outward (inward) Berry-curvature flow, corresponding to Weyl points with positive (negative) chirality.
(b) The slice Chern number $C$ versus $k_{z}$.
(c,d) Numerical energy levels of a finite-sized circular system with different wave vectors $k_z=0.8\pi$ for (c) and $k_z=0.2\pi$ for (d).
Bulk, edge, and anti-corner states are shown in black, red, and blue, respectively. Insets display the cumulative density of states for all edge states in Panel (c) or anti-corner states in Panel (d), where the radius of each colored circle is proportional to the cumulative density $\rho(\mathbf{i})$ at the corresponding lattice site. The other parameters are set as $m=-1$, $B_x=1.5$, $\lambda=0.5$, $B_z=1$, and $R=20a$.}
	\label{fig2}
\end{figure}

In Fig. \ref{fig2}(a), we compute the Berry-curvature distribution of the lower band over
the entire Brillouin zone, using a dense $200^{3}$ momentum grid.
For the numerical calculation we set the parameters \(m = -1\), \(B_x = 1.5\), \(\lambda = 0.5\) (so that \(B_y = \lambda B_x = 0.75\)), and \(B_z = 1\). The curvature exhibits pronounced peaks at the two Weyl points: positive monopoles at \((0, \pm \pi, +k_z^0)\) and negative monopoles at \((0, \pm \pi, -k_z^0)\), with \(k_z^0 \approx 0.42\pi\). This distribution confirms the analytical prediction that Weyl points emerge at high-symmetry points where \(\sin k_x = 0\), \(\sin k_y = 0\), and \(d_z(\mathbf{k}) = 0\).
Integrating \(\boldsymbol{\Omega}(\mathbf{k})\) over each constant-\(k_z\) slice reproduces the slice Chern numbers obtained analytically, namely \(C = 0\) for \(|k_z| < k_z^0\) and \(C = +1\) for \(|k_z| > k_z^0\), as shown in Fig. \ref{fig2}(b). The quantized jump of \(C\) at \(k_z = \pm k_z^0\) directly reflects the topological phase transition driven by the crossing of Weyl points.

The distinct boundary signatures of the two topological phases 
for \(|k_z| < k_z^0\) and for \(|k_z| > k_z^0\)
are further illustrated in Figs. \ref{fig2}(c) and \ref{fig2}(d), which display the numerical energy levels of a finite-sized circular system for fixed \(k_z\). The tight-binding Hamiltonian employed in the calculations is presented in Sec. S-2 of the SM \cite{seesupplemental}.
To construct the circular nanoflake geometry, we consider a square lattice in the $x$-$y$ plane centered at the origin $\left(0,0\right)$, where the position of each site is labeled by its coordinates $\left(p,q\right)$. The sites belonging to the circular nanoflake satisfy the condition $p^2+q^2\le R^2$ with radius $R=20$.
For \(k_z = 0.8\pi\) (corresponding to the QAHI phase with \(C = +1\)), the spectrum shows a set of in-gap states (red dots) that are uniformly distributed along the energy axis. 
To further visualize their spatial distribution, we calculate the cumulative density
$\rho(\mathbf{i})=\sum_{n\in\mathcal{S}}\sum_{\mu=1}^{2}
|\phi_{n,\mu}(\mathbf{i})|^2$ ,
where $\phi_{n,\mu}(\mathbf{i})$ denotes the $\mu$-th component of the $n$-th eigenstate at lattice site $\mathbf{i}$, and $\mathcal{S}$ denotes the selected in-gap states. Physically, $\rho(\mathbf{i})$ represents the accumulated local density of the selected states at each lattice site.
The inset shows the cumulative density of these edge states, confirming their uniform localization along the entire boundary. In contrast, for \(k_z = 0.2\pi\) (corresponding to the anti-HOTI phase with \(C = 0\)), the in-gap states (blue dots) exhibit a different spatial character. While they also appear within the bulk gap, their cumulative density (inset) reveals that they extend along most of the boundary but vanish at specific corners. These are the anti-corner states, the hallmark of the anti-higher-order topology. The comparison between the two panels clearly demonstrates the evolution from conventional chiral edge states in the QAHI phase to the directionally suppressed anti-corner states in the anti-HOTI phase, as the system passes through the Weyl point at \(k_z = k_z^0\).

\begin{figure}
	\centering
	\includegraphics[width=\columnwidth,angle=0]{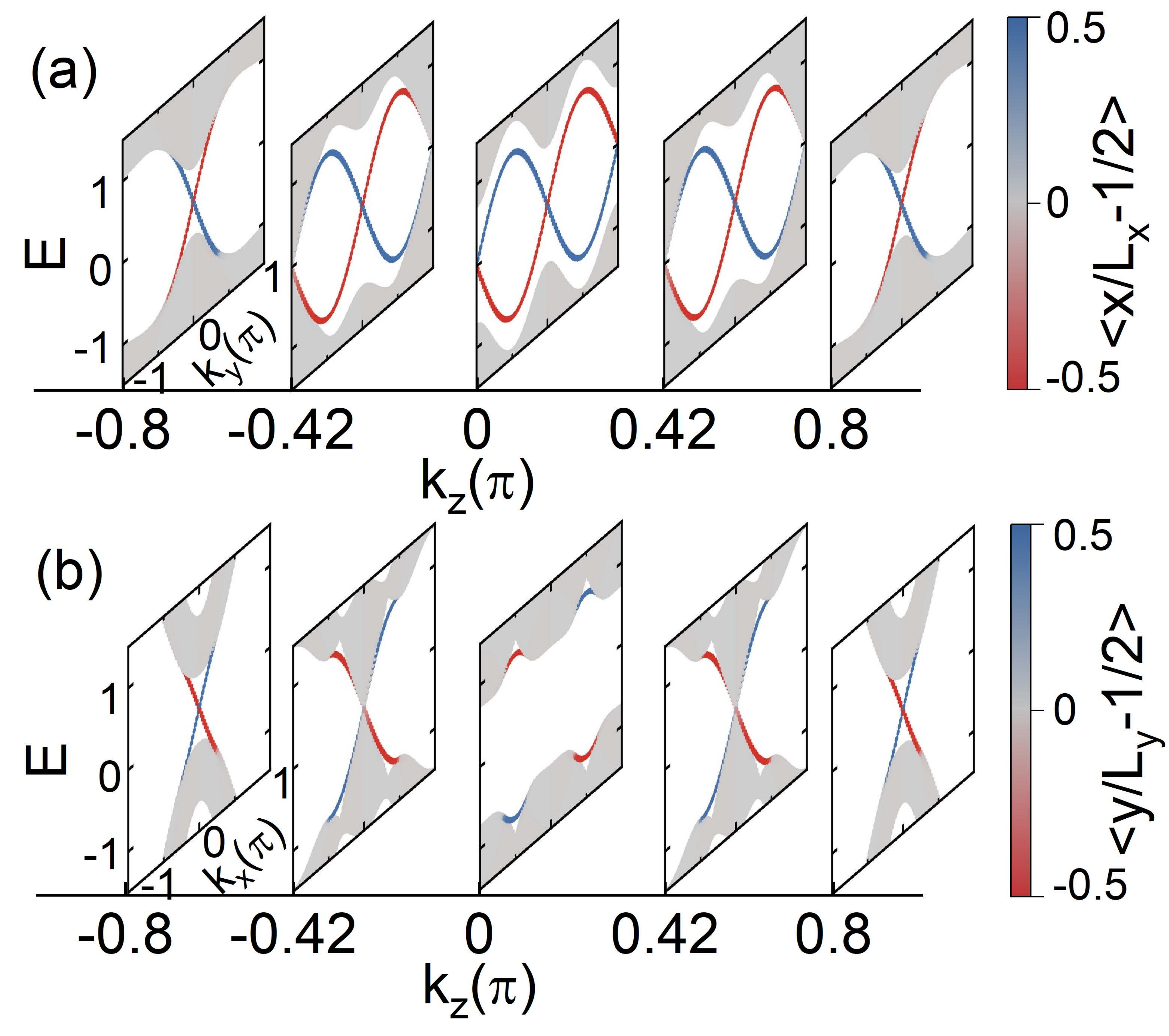}
	\caption{(a) Energy spectra of ribbons versus $k_{y}$, with eigenstates color-coded by their position $\langle x/L_x-1/2\rangle$ to distinguish bulk states and surface (anti-hinge) states. (b) Energy spectra of ribbons versus $k_{x}$, with eigenstates color-coded by $\langle y/L_y-1/2\rangle$ to distinguish bulk states and surface (anti-hinge) states. The width of ribbon is chosen as $L=100$. Other parameters are the same as those in Fig. \ref{fig2}.}
	\label{fig3}
\end{figure}

\emph{Anti-Fermi arcs}.---The definitive fingerprint of an anti-HOWSM is the emergence of "anti-Fermi arcs." Unlike conventional Weyl semimetals where Fermi arcs typically form continuous surface states states connecting Weyl points, the anti-Fermi arcs in anti-HOWSM exhibit a unique subtractive topology: these surface states are selectively suppressed along specific crystalline boundaries while persisting on others.
To visualize this, we calculate the energy spectra of nanoribbons finite in the $x$ and $y$ directions, respectively, as plotted in Fig. \ref{fig3}.
Outside the Weyl points ($|k_z| > k_z^0$, e.g., $k_z = \pm 0.8\pi$), the system behaves as a standard QAHI. As shown in Figs. \ref{fig3}(a) and \ref{fig3}(b) with $k_z = \pm 0.8\pi$, chiral surface states appear on both $x$- and $y$-normal boundaries, confirming a conventional topological phase in these slices.

The signature of the anti-higher-order topology emerges in the region between the Weyl points ($|k_z| < k_z^0$). At $k_z = 0$, the contrast between the two geometries is striking. The ribbon finite along the $x$-direction [Fig. \ref{fig3}(a) with $k_z = 0$] retains robust gapless near $k_y=0$ and $k_y=\pi$. However, crucially, the spectrum for the ribbon finite along the $y$-direction [Fig. \ref{fig3}(b) with $k_z = 0$] becomes completely gapped. 
For ribbons with other orientations, gapless boundary states are generally present, as shown in Secs. S-3 and S-4 of the SM \cite{seesupplemental}.
This vanishing of surface states along the $y$-normal boundaries is not accidental but is the hallmark of the anti-topological phase. It indicates that the "anti" nature of the bulk topology destructively interferes with the boundary modes in specific directions, effectively "erasing" the Fermi arcs on the $y$-surfaces.
Consequently, the Weyl points are connected not by a continuous surface sheet, but by discontinuous anti-Fermi arcs. These arcs exist on all other surfaces, but are interrupted on the $y$-surfaces. This anisotropic feature is completely opposite to the standard HOWSM phase, where topological surface states are preserved along specific directions (i.e., hinge states) and fully suppressed in others.

\begin{figure}
	\centering
	\includegraphics[width=\columnwidth,angle=0]{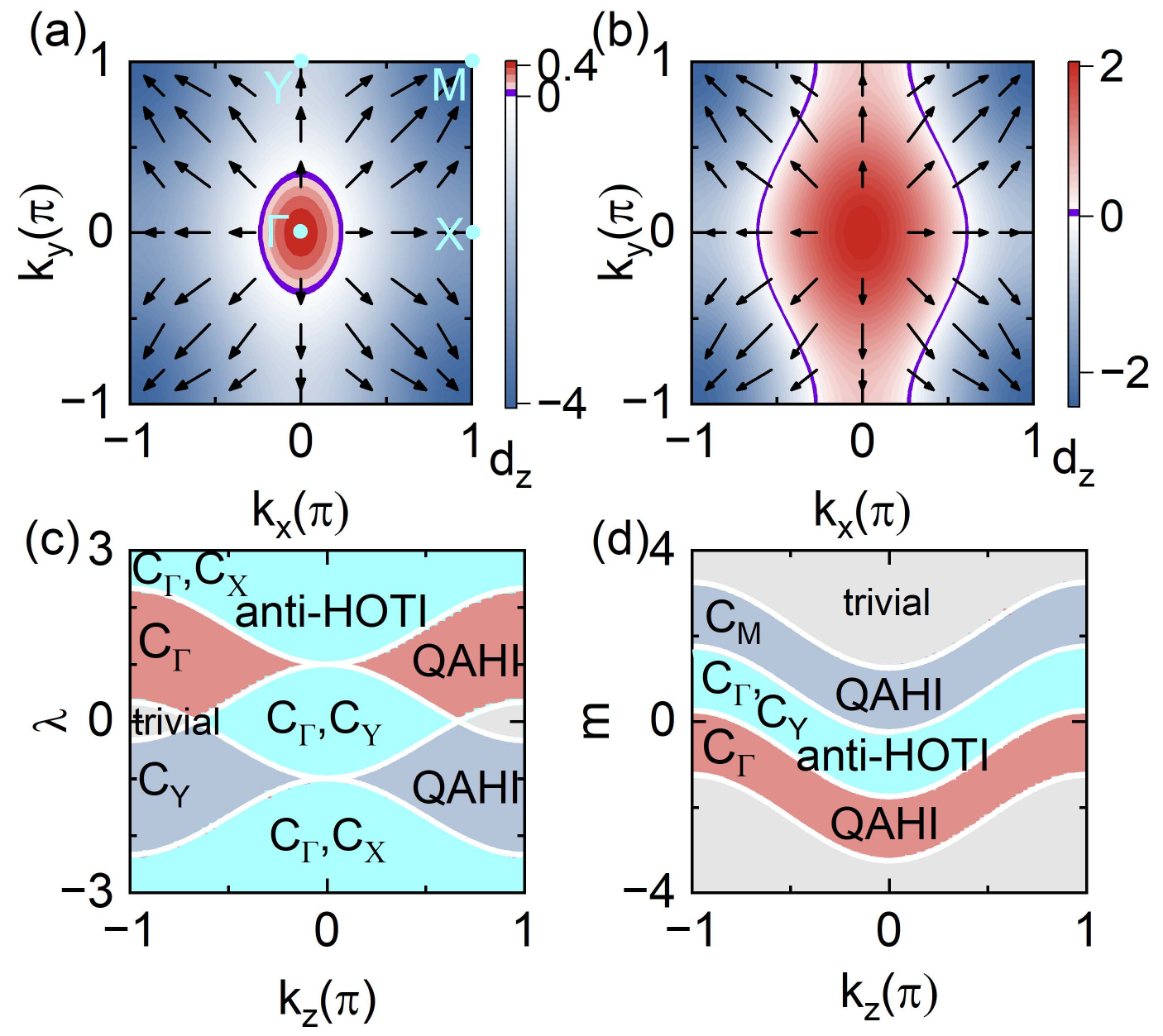}
	\caption{(a,b) Spin-texture maps for representative 2D slices in the QAHI phase at $k_z=0.8\pi$ and the anti-HOTI phase at $k_z=0.2\pi$, respectively.
The color scale represents the out-of-plane component $d_z$, with the purple contour indicating the BIS ($d_z=0$).
Arrows denote the in-plane components $(d_x, d_y)$.
(c,d) Phase diagram as a function of the wave vector $k_z$ and the proportional parameter $\lambda$ or mass parameter $m$. Gray regions correspond to topologically trivial phases. The red (blue) regions indicate QAHI phases, where the BIS encloses a single high-symmetry point with positive (negative) chirality. Cyan regions mark the anti-HOTI phase, where the BIS simultaneously encloses two distinct high-symmetry points with opposite chiralities. White lines indicate the analytically determined boundaries where the bulk gap closes. The other parameters are the same as those in Fig. \ref{fig2}(b).}
	\label{fig4}
\end{figure}

\emph{Topological characterization and phase diagram}.---To classify and characterize the anti-higher-order topological phases identified above, we analyze the distribution of topological charges within fixed $k_z$ slices in our model. 
These charges are distinguished from the monopole charges of 3D Weyl points. The topology is determined by the distribution of topological charges within the band inversion surface (BIS) \cite{zhang_dynamical_2018,zhang_dynamical_2019,wan_helical_2025,wan_interplay_2025}. The BIS is defined as the momentum contour where the mass term of the Hamiltonian vanishes $d_z\left(\mathbf{k}\right)=0$.
The topological charge $\mathcal{C}_\alpha$ is defined as the winding number of the vector field $[d_x(\mathbf{k}), d_y(\mathbf{k})]$ around the singularity $\mathbf{k}_\alpha$ satisfying $d_x(\mathbf{k}_\alpha)=d_y(\mathbf{k}_\alpha)=0$. In our model, these singularities are pinned at the four high-symmetry momenta points $\Gamma=\left(0,0\right), X=\left(\pi,0\right), Y=\left(0,\pi\right)$ and $M=\left(\pi,\pi\right)$, as indicated by four cyan dots in Fig. \ref{fig4}(a).
Recent studies have shown that Chern topological insulators can be further classified by labeling the high-symmetry point enclosed by the BIS \cite{wan_classification_2025}. In this picture, edge states originate from the topological charges enclosed by the BIS, while the enclosed symmetry points determine the momenta at which the edge states appear.
Building on this observation, we define the BIS-enclosed topological charge set
$\mathcal{C}_{BIS}=\left\{\mathcal{C}_\alpha\mid\alpha\in\{\Gamma,X,Y,M\},\alpha\mathrm{\ enclosed\ by\ the\ BIS\ }\right\}$,
which characterizes the distribution and chirality of the topological charges enclosed by BIS.

The spin-texture analysis provides microscopic insight. Figure \ref{fig4}(a) shows the $\mathbf{d}$-vector map for the 2D fixed-$k_z$ slice in the QAHI phase, where the BIS (purple contour) exclusively encloses the $\Gamma$ point. Since the topological charge at the enclosed $\Gamma$ point is $+1$, resulting in the set $\mathcal{C}_{BIS} = \{ \mathcal{C}_\Gamma = +1 \}$. This single component leads to uniform chiral edge states originating from the $\Gamma$ point.
Figure \ref{fig4}(b) displays the anti-HOTI phase, where the BIS encloses both $\Gamma$ and $Y$ points simultaneously. The topological charges at these points are $+1$ and $-1$, respectively, which yields the set $\mathcal{C}_{BIS} = \{ \mathcal{C}_\Gamma = +1, \mathcal{C}_Y = -1 \}$.
The coexistence of these oppositely chiral topological charges gives rise to two sets of counter-propagating edge states that project to different momenta ($\Gamma$ and $Y$). Along specific edge orientations, corresponding to open boundaries along the $y$ direction, these projections overlap in momentum space, leading to the opening of a spectral gap. In contrast, for other boundary orientations, the projections remain separated, and the corresponding boundary states stay gapless. This angle-dependent hybridization mechanism directly explains the formation of anti-corner states and the anti-hinge behavior observed in finite geometries. While the BIS characterization provides an intuitive momentum-space picture of the anti-HOTI phase and naturally explains the projections of the edge states, it does not constitute a symmetry based topological invariant. 
To further quantify the anti-HOTI phase, we also construct the symmetry indicators from the parity eigenvalues at the high-symmetry points, as discussed in Sec.~S-5 of the SM \cite{seesupplemental}.

The phase diagram as a function of $k_z$ and the proportional parameter $\lambda$ (mass parameter $m$) is presented in Fig. \ref{fig4}(c) [\ref{fig4}(d)]. The phase boundaries (white lines) are determined analytically from the bulk gap closing conditions. They consist of four parabolic curves that satisfy Eq.~(\ref{eq4}). Gray regions correspond to trivial insulating phases. Red/blue regions represent first-order QAHI phases. The cyan regions indicate the anti-HOTI phase, in which the band inversion surface simultaneously encloses two distinct high symmetry points carrying opposite topological charges. This configuration underlies the emergence of anti-hinge states and distinguishes the anti-higher-order topological phases from other phases. Importantly, this also explains why the anti-HOTI phase can host remaining boundary states even when the net slice Chern number vanishes. The $C=0$ anti-HOTI phase remains fully consistent with the bulk-boundary correspondence, since the total Chern number determines only the net chirality of the boundary modes, while the momentum-space distribution and projection of the high-symmetry-point topological charges further control the boundary spectral structure.

It is worthy to clarify the relation between the different topological quantities in our work. 
At the 3D level, the Weyl semimetal phase is characterized by the chirality of the Weyl points associated with the monopole structure of the Berry curvature. 
For each fixed $k_z$ slice, the topology is characterized by the slice Chern number $C$. 
However, since the anti-HOTI phase satisfies $C=0$, the slice Chern number alone is insufficient to distinguish it from a trivial insulating phase. 
To further characterize the anti-HOTI phase, we therefore employ both the BIS winding charge picture and the symmetry indicators constructed from the parity eigenvalues at the high-symmetry points, which provide complementary characterizations of the topology for the fixed $k_z$ slices. 
Together, these different topological quantities characterize distinct aspects of the topology in the anti-HOWSM system.

\emph{Discussion}.---In summary, we have identified an anti-higher-order topological phase in a 3D Weyl semimetal framework. Distinct from conventional higher-order topology, this phase is characterized by a directional suppression of boundary states rather than protected lower-dimensional modes. By analyzing both bulk topological invariants and boundary spectra, we show that the anti-HOTI phase emerges in the momentum region between a pair of Weyl points and is characterized by a BIS enclosing multiple high-symmetry points with opposite topological charges. This configuration leads to a destructive interference of boundary projections, resulting in anti-hinge and anti-Fermi arcs responses, where the systematic absence of lower-dimensional boundary states, constitutes the defining fingerprint of the anti-higher-order topological phase.

Weyl semimetals stabilized by broken inversion or time-reversal symmetry have been experimentally realized in several electronic materials \cite{armitage_weyl_2018, lv_experimental_2015, jiang_signature_2017}. Building on these established platforms, introducing additional anisotropies, altermagnetism \cite{smejkal_beyond_2022}, or strain \cite{webster_straintunable_2018} may deform the band-inversion surface at fixed momentum slices, offering a natural route toward enclosing topological charges of opposite chirality and realizing the anti-higher-order Weyl phase proposed here. Beyond electronic systems, artificial platforms such as photonic, acoustic, and electrical-circuit lattices provide exceptional control over symmetry and coupling geometry, and therefore represent especially promising settings for exploring anti-higher-order boundary responses \cite{wei_higherorder_2021, pan_real_2023,  zheng_topolectrical_2022}. 
In these artificial platforms, the clearest observable signature is the boundary-orientation-dependent response, where boundary states remain visible along most boundaries but disappear on selected corners or hinges. Experimentally, such behavior can be identified from boundary-resolved spectra, spatial field distributions, surface equifrequency contours in photonic and acoustic lattices~\cite{lu_topological_2014,ozawa_topological_2019}, impedance spectra in topolectrical circuits~\cite{imhof_topolectrical_2018,lee_topolectrical_2018}, and momentum-resolved angle-resolved photoemission spectroscopy measurements in electronic systems~\cite{xu_discovery_2015,belopolski_criteria_2016}. This distinguishes the proposed anti-Fermi arc from both the ordinary surface Fermi arc in a Weyl semimetal and the hinge Fermi arc in a conventional HOWSM. A more detailed comparison between the conventional Weyl semimetal, HOWSM, and anti-HOWSM is provided in Sec.~S-6 of the SM \cite{seesupplemental}.

The concept of anti-higher-order topology introduced here extends the conventional bulk–boundary correspondence by emphasizing the controlled suppression of boundary states. It provides a general framework applicable to a broad class of topological semimetals and insulators and opens new avenues for engineering directional boundary responses in topological materials. The topology protected suppression of boundary states may further provide possible functionalities for directional transport blocking, boundary-state filtering, and programmable wave guiding. In particular, boundary states can exhibit robust absence along selected boundary orientations due to the topological projection and hybridization mechanism rather than local defects or accidental scattering. Therefore, such boundary transport interruptions are expected to remain stable against moderate disorder and impurity scattering.

\emph{Acknowledgements}.---This work was financially supported by the National Key Research and Development Program of China (Grant No. 2024YFA1409002), the National Natural Science Foundation of China (Grants No. 12374034, No. 124B2069, and No. 12447147), the Quantum Science and Technology-National Science and Technology Major Project (Grant No. 2021ZD0302403), and the China Postdoctoral Science Foundation (Grant No. 2024M760070). We also acknowledge the High-performance Computing Platform of Peking University for providing computational resources.\\

\bibliography{AHOTSM}

\end{document}